\documentclass[10pt,conference]{IEEEtran}
\usepackage[utf8]{inputenc}
\usepackage{graphicx}
\usepackage[bookmarks=false]{hyperref}
\usepackage{listings}
\usepackage{courier}

\title{Kotless: a Serverless Framework for Kotlin}
\author{
\IEEEauthorblockN{Vladislav Tankov}
\IEEEauthorblockA{JetBrains\\
ITMO University\\
Email: vladislav.tankov@jetbrains.com}
\and
\IEEEauthorblockN{Yaroslav Golubev}
\IEEEauthorblockA{JetBrains Research\\
ITMO University\\
Email: golubev@itmo.ru}
\and
\IEEEauthorblockN{Timofey Bryksin}
\IEEEauthorblockA{JetBrains Research\\
Saint-Petersburg State University\\
Email: t.bryksin@spbu.ru}
}

\begin{document}

\maketitle

\begin{abstract}
Recent trends in Web development demonstrate an increased interest in serverless applications, i.e. applications that utilize computational resources provided by cloud services on demand instead of requiring traditional server management. This approach enables better resource management while being scalable, reliable, and cost-effective. However, it comes with a number of organizational and technical difficulties which stem from the interaction between the application and the cloud infrastructure, for example, having to set up a recurring task of reuploading updated files. In this paper, we present Kotless --- a Kotlin Serverless Framework. Kotless is a cloud-agnostic toolkit that solves these problems by interweaving the deployed application into the cloud infrastructure and automatically generating the necessary deployment code. This relieves developers from having to spend their time integrating and managing their applications instead of developing them. Kotless has proven its capabilities and has been used to develop several serverless applications already in production. Its source code is available at \url{https://github.com/JetBrains/kotless}, a tool demo can be found at \url{https://www.youtube.com/watch?v=IMSakPNl3TY}.

\end{abstract}

\section{Introduction}

The term \textit{serverless computing} refers to the concept of building and running applications that do not require traditional server management~\cite{CNCF_SERVERLESS_OVERVIEW}. Such execution model implies that applications are uploaded to a cloud service provider's infrastructure bundled as one or several functions. Then they are executed on demand, automatically scaled if needed, and billed according to the exact time these applications have been running. This approach liberates developers from the burden of maintaining their own hardware infrastructure and allows to focus more on software~\cite{SERVERLESS_OVERVIEW}.

Unfortunately, developers cannot simply \textit{upload} the code to a cloud service provider, they have to \textit{deploy} it. To support that, the application should be implemented in an event-driven way, providing certain functions which are triggered by internal cloud events or incoming HTTP requests~\cite{SERVERLESS_EVENT_DRIVEN}. Thus, the deployment process of a serverless application requires defining a list of specific events that the application uses and endpoints it provides. For all major cloud service providers, this is done using custom deployment domain-specific languages, or DSLs. Despite the fact that almost all interaction between the cloud infrastructure and the application is defined within the application's code, it is still necessary to write a lot of boilerplate code, which often redefines constructs that are already present in the deployment DSL code. Moreover, the total amount of deployment code in a complex application may even be comparable to the size of the application itself.

In this paper, we present Kotless --- a serverless framework for Kotlin applications, which is open-source and available on GitHub.\footnote{ \url{https://github.com/JetBrains/kotless}} The idea of Kotless is to provide a cloud-agnostic DSL that unites the cloud service provider's SDK and the deployment DSL. Using Kotless, a developer simply writes an application in terms of event-driven architecture and the deployment code is generated automatically straight from the source code.

Our approach eliminates the need for a specific deployment DSL and makes serverless computing comprehensible to anyone familiar with event-based architectures. For example, a developer only needs to be familiar with JAX-RS\footnote{JAX-RS is a Java API for RESTful web services, described in JSR311: \url{https://jcp.org/en/jsr/detail?id=311}}-like annotations to create and deploy a REST API application based on Kotless.

\section{Overview}\label{overview}

The main idea of Kotless is to interweave the deployed application into the deployment infrastructure --- to create a framework that allows to use certain cloud services in the application's code and then automatically create an appropriate infrastructure tailored for this application.

To demonstrate this idea let us consider an example of a very basic web application that needs to host static files and provide access to them via HTTP. Normally, its developers would ask the IT department to configure an HTTP server for them or to create an Amazon S3 bucket. Then a recurring task is to upload new files after each new deployment of this application --- quite a lot of work overall. With Kotless, this task could be described with the snippet of Kotlin code shown in Listing~\ref{lst:static-get}. 

\vspace{0.2cm}
\begin{lstlisting}[language=Java, caption={A fragment of a web application written with Kotless}, captionpos=b, label={lst:static-get}]
@StaticGet("/style.css", MimeType.CSS)
val style = File("css/style.css")
\end{lstlisting}
\vspace{0.2cm}

In this example, the $StaticGet$ annotation represents a cloud service request for a static resource accessible via HTTP. Kotless deployment tools find this request and make sure that the cloud infrastructure stores the necessary resource. And moreover, this resource (a CSS file in this example) will be re-deployed when it changes. 

Other cloud services (dynamic HTTP handlers, permissions, etc.) are handled in a similar fashion. In practice, this approach is extensible to any available cloud service provider or even self-hosted solutions like OpenShift. 

\section{Architecture and tool details}

Kotless is a Kotlin framework that consists of two major parts:
\begin{itemize}
    \item Kotless DSL --- a client-side DSL, distributed in the form of a library;
    \item Kotless plugin --- a deployment tool, distributed in the form of a Gradle plugin.
\end{itemize}

It is worth noting that Kotless does not interact with the cloud provider itself, instead, it generates Terraform code. Terraform\footnote{Terraform is a HashiCorp deployment tool: \url{https://www.terraform.io}} is one of the popular deployment DSLs and is described in more detail in Section~\ref{sec:alternatives}. In this work, it is used as a kind of a \textit{cloud bytecode}: the Terraform code is generated by Kotless to perform the actual deployment of the application. 

\begin{figure}
  \centering
  \includegraphics[width=3.3in]{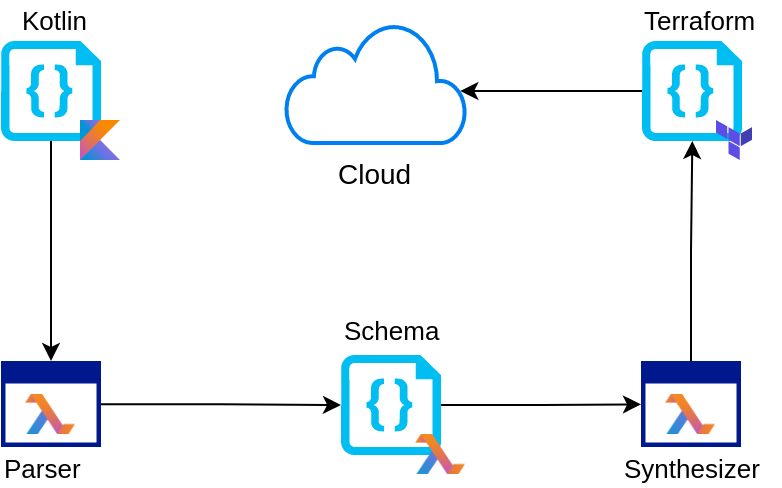}
  \caption{Kotless pipeline}
  \label{pipeline}
\end{figure}

Now, let us take a look at the entire pipeline to see how the source code written with Kotless becomes a fully deployed serverless web application (Figure~\ref{pipeline}) and then overview the main capabilities of Kotless DSL and Kotless plugin internals.

\subsection{Kotless pipeline}

First of all, Kotless DSL is used to define HTTP endpoints and other resources that exist in the cloud (e.g. database tables, object storage, etc.). The DSL is cloud-agnostic, meaning that if the developer decides to switch from one cloud service provider to another, no changes in the application code will be required to re-deploy the application (of course, only in case the application is not integrated with cloud-specific services used outside of Kotless abstractions).

Before the first deployment, the developer sets up credentials to access the cloud provider, as well as some configuration parameters that define the storage that Kotless might utilize. Once that is over with, the developer executes the deployment task of the Kotless Gradle plugin. This task prepares the environment (downloads Terraform, allocates temporary storage, etc.) and launches Kotless DSL Parser.

Kotless DSL Parser is an library capable of parsing Kotlin code and extracting declarations (for example, annotations in Listing~\ref{lst:static-get}) of Kotless DSL. It is based on the embeddable version of Kotlin compiler and can be used not only from Gradle, but from any Java application (for example, from an integrated development environment, such as IntelliJ IDEA). It parses the application code and generates a Kotless Schema object.

Kotless Schema is a cloud-agnostic serverless model that defines available serverless resources, such as dynamic and static HTTP routes, serverless functions, and others. Each Kotless Schema object represents resources that should be deployed for this particular application. This object is then passed to Kotless Terraform Synthesizer, which uses it to generate pretty-printed Terraform code. The generation of pretty-printed code includes not only formatting the code but also its separation into groups by types of services and sorting them within each group according to resource dependencies.

After the Terraform code is generated, Kotless plugin prepares a distributable jar file of the application with all the necessary dependencies and deploys it with generated deployment code. However, the work of Kotless does not stop on the developer's side. The framework also handles requests to the application from the cloud in runtime and dispatches them in accordance with the obtained Schema description.

\subsection{Kotless DSL}
Kotless DSL is a set of interfaces for various subsystems that helps developers build serverless applications.
It is used both during the deployment stage and in runtime, which means that certain DSL constructs will be firstly used to allocate all the necessary cloud services, and later the same constructs will handle the interaction between the application and the allocated services. For example, $Get$ annotation is used to generate HTTP routes during the deployment stage and to dispatch requests in runtime. In order for that to work, the deployment tool must be integrated into the application, which was one of the main reasons for creating Kotless in the first place. 

The DSL consists of three major components.

\subsubsection{HTTP API}

This component includes annotations for creating HTTP routes, deploying static files, extensible serialization, and intercepting HTTP requests. HTTP API provides everything that a standard library of a Web framework is expected to provide. However, one of the major elements of this API is a \textit{Dynamic route} construct. Dynamic routes are Kotlin functions annotated with $Get$ or $Post$, which create new HTTP handlers with an appropriate HTTP method. In case these functions have parameters or return a result of a primitive type, the API performs automatic deserialization of such values.
To extend this feature to non-primitive types the programmer has to implement a conversion service object.

An example of a dynamic $Get$ route is shown in Listing~\ref{lst:dynamic-get}. When the root page is requested, the application returns the "Hello world!" string. 
In case of Amazon Web Services (AWS), deploying this snippet will lead to creating an API Gateway,\footnote{API Gateway is a API service managed by AWS: \url{https://aws.amazon.com/api-gateway}} an AWS Lambda\footnote{AWS Lambda is a serverless function implementation by AWS: \url{https://aws.amazon.com/lambda}} instance, and the integration between them~\cite{AWS_SERVERLESS_ARCHITECTURE}. The resulting application will be globally accessible via a domain name created for the API Gateway.

\vspace{0.2cm}
\begin{lstlisting}[showstringspaces=false,language=Java, caption={Dynamic route declaration}, captionpos=b, label={lst:dynamic-get}]
@Get("/")
fun root(): String {
    return "Hello world!"
}
\end{lstlisting}
\vspace{0.2cm}

\subsubsection{Lifecycle API}

This component includes various interfaces to control and extend the lifecycle of serverless functions, supporting initialization and warming procedures.

\textit{Initialization} is a well-known operation that is supported by almost any existing web framework, whereas \textit{warming} refers to the lifecycle of serverless functions and their problem of a cold start~\cite{SERVERLESS_TRENDS}. A cold start occurs when a serverless function is invoked for the first time and the code packages are created and initialized. Repeated calls to this function will result in lower response time since the function will already be loaded into the server's memory (such functions are called \textit{warm pool}). Cloud service providers implement their own heuristics regarding which functions should stay in this warm pool and for how long, but most of them are based on how many times a function has been called during a specific time period. If a function is not in high demand, it is unloaded and will go through a cold start once again when called next time.

To deal with this problem Kotless introduces a scheduled event: every N minutes an event triggers a warming sequence inside a Kotless-based application. 
This feature does not bring a lot of extra costs but significantly lowers the application's latency by mitigating the likelihood of a cold start.

\subsubsection{Permissions API}
This component provides a declarative way to bind access to Kotlin code entities with permissions on the cloud service provider's side. These annotations can be used on classes, Kotlin static objects, functions, and properties. Each annotation states that the access to this particular element in Kotlin code requires permissions to the cloud provider resource stated in the annotation. This means that each access to the annotated entity will grant a certain cloud service permission to this code.
Listing~\ref{lst:permission-tailor} features an example object that is granted Read and Write permissions to a DynamoDB\footnote{DynamoDB is a NoSQL database by AWS: \url{https://aws.amazon.com/dynamodb}} table \textit{id}.

\vspace{0.2cm}
\begin{lstlisting}[language=Java, caption={Permissions API usage}, captionpos=b, label={lst:permission-tailor}]
@DynamoDBTable("id", ReadWrite)
object Storage {
    val table = DynamoTable("id")
}
\end{lstlisting}
\vspace{0.2cm}

\subsection{Kotless plugin}

Kotless plugin is a key component of the Kotless framework. It connects the abstract representation of a serverless application and the process of its actual deployment. The plugin executes the aforementioned pipeline: it runs the parser and the synthesizer, prepares the environment for the deployment, and performs the deployment itself. Currently, Kotless plugin is distributed as a Gradle plugin. However, we are considering implementing it for other build tools as well. 

\section{Evaluation}
\subsection{Typical use cases}

There are two major application types our framework is best suitable for. First of them are low-load infrastructural applications. For example, a simple adapter service that transforms incoming events into some other output events and transfers them to another service. As mentioned in Section~\ref{overview}, deployment of such an application requires a physical server, which the company has to pay for, even if it is not fully loaded. And moreover, maintaining it involves both software engineering and IT departments, which increases its cost. With Kotless, most of this maintenance routine is handled by our framework. The application is serverless and is hosted in the cloud, so there is no need for a dedicated server and the company will be billed according to the actual load --- if there are no incoming requests, the application costs nothing.

Another common usage scenario involves applications with strict scaling requirements~\cite{SERVERLESS_ECONOMICS}. For example, in AWS, serverless applications are allowed to scale up to 1000 AWS Lambda instances running in parallel, which means that at peak load the application might scale up to 3 Tb of RAM (3 Gb per Lambda instance) and 2000 CPUs (2 CPUs per Lambda instance). With an average request processing time of approximately 200 ms, it provides a processing bandwidth of 5000 requests per second. If the load is then lowered, the application will immediately scale down to 1 Lambda instance, ensuring effective resource management.

\subsection{Case study}

Kotless has already been used internally in JetBrains to develop several infrastructural services. One of them is TrackGenie --- a web service that transforms OpsGenie\footnote{OpsGenie is an incident management platform developed by Atlassian: \url{https://www.opsgenie.com}} alerts into YouTrack\footnote{YouTrack is an issue tracker developed by JetBrains: \url{https://www.jetbrains.com/youtrack}} issues describing the incident. It handles tens of alerts every day and has proven its reliability over the course of several months. It has also brought the price of the application use in the AWS cloud from \$55 down to \$1.85 per month. Besides, TrackGenie is capable of fast scaling: when OpsGenie was temporarily misconfigured one time, it loaded TrackGenie with more than 300 requests per second and the application handled this load successfully.

TrackGenie was developed by a single software developer not experienced in cloud technologies or Terraform in just three days. Without Kotless, it would be nearly impossible --- the tool generated more than a thousand lines of complex Terraform code for TrackGenie deployment. It only took 15 Kotless annotations to define this behavior.

\section{Existing alternatives}\label{sec:alternatives}

There are plenty of deployment tools available on the market today, let us take a closer look at the most popular ones and see how they compare to Kotless.

Terraform is a cloud-agnostic resource provisioning tool developed by HashiCorp~\cite{TERRAFORM_BOOK}. It uses HashiCorp configuration language (HCL) to declare resources and has plenty of so-called \textit{providers} that allow to integrate Terraform into different cloud infrastructures. The main difference between Kotless and Terraform is that Kotless is not a separate tool, it is integrated into an application, while Terraform is merely a deployment DSL. As mentioned previously, Terraform is used as a kind of a \textit{cloud bytecode} in our framework because of its maturity and cloud-agnosticism. 

There are three products developed by Amazon for resource provisioning in AWS: CloudFormation, SAM and CDK. CloudFormation\footnote{CloudFormation's official website: \url{https://aws.amazon.com/cloudformation}} uses a declarative DSL and is very similar to Terraform, however, it has an important downside of an AWS vendor lock-in~\cite{AWS_BOOK}. SAM\footnote{SAM's official website: \url{https://aws.amazon.com/serverless/sam}} (Serverless Application Model) is a framework for building serverless applications for AWS. Basically, it is a subset of CloudFormation with a slightly different syntax and a large number of available tools to make the development of a serverless application easier for developers. CDK\footnote{CDK's documentation website: \url{https://docs.aws.amazon.com/cdk/latest/guide/home.html}} (Cloud Development Kit) is a new tool that was presented by AWS after the work on Kotless has already started. It is a library that provides a set of interfaces to create various AWS services. While CDK is the closest alternative to Kotless, it operates differently: CDK offers a procedural way of creating deployment pipelines, but it does not integrate the deployment process into an application and requires creating a separate \textit{deployment application}. In future work, we are strongly considering using CDK as a part of the Kotless client-side interface.

Another very popular solution for serverless deployment is Serverless.com.\footnote{Serverless.com official website: \url{https://serverless.com}} It uses a YAML\footnote{YAML is a human-friendly data serialization standard, its official website: \url{https://yaml.org}}-based declarative DSL and is a close analog of SAM, however, unlike SAM, it supports multiple cloud service providers (AWS, Azure, Google Cloud, etc.) and a lot more services within each of them. Despite having a rich API, a separate DSL code is used to deploy an existing application.

Thus, none of the existing alternatives support integration with the application to the extent Kotless does. Despite a high degree of flexibility that Terraform or CloudFormation provide, Kotless has a significantly lower learning curve than these instruments.

\section{Conclusion and Future work}

In this paper, we present Kotless --- a production-ready framework for the development of web services. Based on the internal use at JetBrains, it has already proven itself to be a reliable, scalable and inexpensive solution for a vast variety of web applications in a production environment. 

The framework is currently evolving in the following directions. First of all, we are working on the implementation of Construction API, which enables integration with CDK or any other similar toolkit. This allows users to create and modify the infrastructure in runtime, for example, set up tables in DynamoDB or add new HTTP endpoints. As a part of this API, we plan to provide users with a possibility to define the infrastructure at compile time as well. That allows to avoid granting too broad permissions to serverless functions, which increases the security of the developed applications.

Another direction of future work is Event API, containing support of auxiliary cloud events. For example, events that are triggered when new records are added to an SQS\footnote{SQS is a service for managed message queues by AWS, official website: \url{https://aws.amazon.com/sqs}} queue. This API would also allow users to set up scheduled tasks.

Our long-term goal is to provide support for a wide variety of cloud service providers, in particular, Azure and Google Cloud. Despite the fact that the architecture of Kotless is already cloud-agnostic in its foundation, it still takes time to implement Kotless abstractions in terms of particular cloud service providers API. 

We believe that tighter integration of the cloud infrastructure and the application is the future of web services and it requires creating brand new tools suitable for developers: simple, accessible, and concise.

\bibliography{cites} 
\bibliographystyle{ieeetr}

\end{document}